\documentclass[11pt]{article}
\usepackage{jheppub}

\usepackage{empheq}
\usepackage{bbm}

\newcommand{\be}{\begin{equation}}
\newcommand{\ee}{\end{equation}}
\newcommand{\bea}{\begin{eqnarray}}
\newcommand{\eea}{\end{eqnarray}}

\def\Om{{\mathcal{O}}}
\def\Cm{{\mathcal{C}}}
\def\Em{{\mathcal{E}}}
\def\Am{{\mathcal{A}}}
\def\Gm{{\mathcal{G}}}
\def\Bm{{\mathcal{B}}}
\def\Nm{{\mathcal{N}}}

\def\Jm{{\mathcal{J}}}
\def\Im{{\mathcal{I}}}
\def\Km{{\mathcal{K}}}
\def\Qm{{\mathcal{Q}}}

\def\Pm{{\mathcal{P}}}
\def\Dm{{\mathcal{D}}}
\def\Rm{{\mathcal{R}}}
\def\Sm{{\mathcal{S}}}
\def\Mm{{\mathcal{M}}}

\def\a{\alpha}
\def\ad{\dot{\alpha}}
\def\b{\beta}
\def\bd{\dot{\beta}}
\def\c{\gamma}
\def\cd{\dot{\gamma}}
\def\th{\theta}
\def\thb{\bar{\theta}}

\def\Xbf{{\mathbf{X}}}

\def\veps{\varepsilon}
\def\ph{\phantom}
\def\pd{\partial}

\def\nn{\nonumber}

\def\be{\begin{equation}}
\def\ee{\end{equation}}
\def\bea{\begin{eqnarray}}
\def\eea{\end{eqnarray}}

\title{\boldmath Stress-tensor OPE in $\Nm=2$ Superconformal Theories}

\author[a]{Pedro Liendo,}
\author[a,b]{Israel Ram\'irez,}
\author[c]{Jihye Seo.}

\affiliation[a]{IMIP, Humboldt-Universit{\"a}t zu Berlin, IRIS Adlershof, \\Zum Gro{\ss}en Windkanal 6, 12489 Berlin, Germany}
\affiliation[b]{Departamento de F\'isica, Universidad T\'ecnica Federico Santa Mar\'ia, \\
Casilla 110-V, Valpara\'iso, Chile}
\affiliation[c]{DESY Hamburg, Theory Group,\\ Notkestrasse 85, D–22607 Hamburg, Germany}

\emailAdd{pliendo@physik.hu-berlin.de}
\emailAdd{israel.ramirez.12@sansano.usm.cl}
\emailAdd{jihye.seo@desy.de}

\abstract{We carry out a detailed superspace analysis of the OPE of two $\Nm=2$ stress-tensor multiplets. Knowledge of the multiplets appearing in the expansion, together with the two-dimensional chiral algebra description of $\Nm=2$ SCFTs, imply an analytic bound on the central charge $c$. This bound is valid for any $\Nm=2$ SCFT regardless of its matter content and flavor symmetries, and is saturated by the simplest Argyres-Douglas fixed point. We also present a partial conformal block analysis for the scalar superconformal primary of the multiplet.}

\arxivnumber{1509.00033}

\begin{document}
\maketitle
\flushbottom



\section{Introduction}

Four-dimensional superconformal field theories (SCFTs) with $\Nm=2$ supersymmetry play a prominent role in theoretical physics. Originally studied using standard field theoretic tools, by building Lagrangians out of fundamental fields with appropriately chosen matter content, the list of theories has grown considerable in recent years \cite{Gaiotto:2009we,Gaiotto:2009hg}, and now there seems to be an extensive library of $\Nm=2$ systems, related by an intricate web of dualities. Having found such an ample catalog, there has been a shift in perspective, instead of analyzing specific models one by one, it seems more natural to ask whether a classification program is possible. Efforts in this direction include a classification of Lagrangian models \cite{Bhardwaj:2013qia}, a procedure for classifying class $\Sm$ theories \cite{Chacaltana:2010ks}, and a systematic analysis of Coulomb branch geometries \cite{Argyres:2015ffa,Xie:2015rpa}.

Among the most important tools for constraining the space of CFTs is the conformal bootstrap approach \cite{Polyakov:1974gs,Ferrara:1973yt,Mack:1975jr}. Originally very successful in two dimensions, where the conformal algebra is enhanced to the infinite dimensional Virasoro algebra, it has seen renewed interest in light of the work of \cite{Rattazzi:2008pe} where, starting from basic principles like crossing symmetry and unitarity, numerical techniques were developed that allow to obtain rigorous bounds on several CFT quantities. Influenced by this revival of the bootstrap philosophy, the $\Nm=2$ superconformal bootstrap program was initiated in \cite{Beem:2013sza,Beem:2014zpa}, with the goal of serving as an organizing principle, relying only on the operator algebra of a theory, as defined by the OPE.

The $\Nm=2$ superconformal bootstrap program can be thought of as a two-step process. First, it was observed in \cite{Beem:2013sza} that any $\Nm=2$ SCFT contains a protected subsector of observables described by a two-dimensional chiral algebra. In order to bootstrap a full-fledged SCFT, one must first have an understanding of the operators described by the chiral algebra. Once this is achieved, the second step entails tackling the harder task of bootstrapping the full theory, in particular, unprotected operators with unconstrained conformal dimensions. This second step was explored in \cite{Beem:2014zpa} using the numerical techniques of \cite{Rattazzi:2008pe}, and bounds were obtained by looking at four-point correlators of several superconformal multiplets. Though a comprehensive effort toward bootstrapping the landscape of $\Nm=2$ theories, there was an important omission, the multiplet in which the stress-tensor sits was absent from the analysis. The universal nature of the stress-tensor makes it a natural target for bootstrap studies, and the reason it was not included in \cite{Beem:2014zpa} was technical: the requisite crossing symmetry equation is not known. 

Let us be a bit more specific. The conserved stress-tensor of an $\Nm=2$ theory sits in a multiplet that can be represented by a superfield $\Jm$ with a schematic $\th$-expansion,
\be 
\Jm(x,\th,\thb)\vert = J(x)\, , 
\qquad \Jm(x,\th,\thb)\vert_{\th \thb} = J^{ij}_{\mu}(x)\, ,
\qquad \Jm(x,\th,\thb)\vert_{\th^2 \thb^2} = T_{\mu \nu}(x)\, .
\ee
$J$ is a scalar superconformal primary of dimension two, $ J^{ij}_{\mu}$ is the conserved $SU(2)_R \times U(1)_r$ $R$-symmetry current, and $T_{\mu \nu}$ is the stress-tensor. Correlators of this multiplet also contain information about two fundamental quantities present in any four-dimensional CFT, the $a$ and $c$ anomaly coefficients. These can be defined as the anomalous trace of the stress-tensor when the theory is considered in a curved background.

Our first goal will be to obtain the supersymmetric selection rules
\be 
\Jm \times \Jm \sim \ldots\, ,
\ee
namely, the $\Nm=2$ multiplets that are allowed to appear in the super OPE of two stress-tensor multiplets. This result will be relevant for both the two-dimensional chiral algebra description and the numerical bounds program.

On the chiral algebra side, as observed in \cite{Beem:2013sza}, the two-dimensional stress-tensor can be associated to the four-dimensional $SU(2)_R$ current. In particular, correlators of the four-dimensional current have a solvable truncation described by correlators of the two-dimensional holomorphic stress-tensor. The holomorphic correlator only depends on the central charge $c$ and, as we will see in this work, unitarity of the four-dimensional theory implies an analytic bound on $c$. The $a$ anomaly coefficient plays no role in the chiral algebra construction.

On the numerical side, the super OPE selection rules are the first step toward writing the crossing symmetry equation for the stress-tensor multiplet. To have a better understanding of how this can be accomplished, let us recall how the numerical bootstrap is implemented. The starting point is the four-point function of a real scalar operator $\phi$. This correlator can be expanded using  a  \textit{conformal block} expansion, where each conformal block captures the contribution of a specific conformal family appearing in the $\phi \times \phi$ OPE. Explicit expressions for scalar conformal blocks were obtained in \cite{Dolan:2000ut,Dolan:2003hv}. Having obtained such an expansion, using the restrictions imposed by crossing symmetry and unitarity, it is possible to obtain numerical bounds on scaling dimensions and three-point couplings. In $\Nm=2$ theories, the highest weight of the stress-tensor multiplet is a scalar of dimension $\Delta_{J}=2$, and is therefore well suited for the numerical bootstrap program. Because of supersymmetry, several conformal families are related by the action of supercharges, and this implies that a finite number of conformal blocks appearing in a correlator can be grouped together in a \textit{superconformal block}, which encodes the contribution of the corresponding superconformal family. We can now state more precisely why the stress-tensor correlator was not included in \cite{Beem:2014zpa}: the superconformal block expansion of the $\langle J J J J \rangle$ correlator has not been worked out. To fill this gap in the $\Nm=2$ literature was one of the motivations for this work.

Conformal and superconformal block expansions are a common obstacle in any attempt to write bootstrap equations. In the bosonic case, things get very complicated when one considers operators in non-trivial Lorentz representations. With supersymmetry, many complications arise for correlators of generic multiplets. There is no unifying framework and a wide variety of approaches have been tried with varying degrees of success \cite{Costa:2011mg,Dolan:2011dv,Costa:2011dw,SimmonsDuffin:2012uy,Dymarsky:2013wla,Hogervorst:2013kva,Fitzpatrick:2014oza,
Khandker:2014mpa,Elkhidir:2014woa,Costa:2014rya,Echeverri:2015rwa,Bissi:2015qoa,Doobary:2015gia,Rejon-Barrera:2015bpa}. The full superconformal block expression for the $J$ correlator is still elusive, and it is not clear which of all the methods available in the literature is the most efficient. Nevertheless, our calculation encodes the allowed $\Nm=2$ multiplets that contribute to the expansion, 
which is the first step toward writing the crossing symmetry equation.

The outline of the paper is as follows. In section 2 we review the conformal algebra and its shortening conditions. Section 3 presents a detailed superspace analysis that allows us to write the super OPE selection rules for two stress-tensor multiplets. In section 4 we use our selection rules together with the two-dimensional chiral algebra construction to obtain an analytic bound on $c$. This bound is valid for any $\Nm=2$ superconformal theory regardless of its matter content and flavor symmetries. In section 5 we present a partial analysis of the superconformal block expansion of the $J$ correlator.

\section{Preliminaries}

The $\Nm=2$ superconformal algebra is the algebra of the supergroup $SU(2,2|2)$. It contains the conformal algebra  $SU(2,2) \sim SO(4,2)$ with generators $\{\Pm_{\a \ad},\, \Km^{\ad \a},\, \Mm_{\a}^{\ph{\a} \b},\, \bar{\Mm}^{\ad}_{\ph{\ad}\bd},\, D\}$, where $\a=\pm$ and $\ad=\dot{\pm}$ are Lorentz indices, and an $SU(2)_R \times U(1)_r$ $R$-symmetry algebra with generators $\{\Rm_{\ph{i}j}^{i}, \, r \}$, where $i=1,2$ are $SU(2)_R$ indices. In addition to the bosonic generators there are fermionic supercharges, the Poincar\'e and conformal supercharges, $\{\Qm^i_{\a},\bar{\Qm}_{i\, \ad}, \Sm_{i}^{\a},\bar{\Sm}^{i\, \ad}\}$.

A general supermultiplet of $SU(2,2|2)$ contains a highest weight or superconformal primary with quantum number $(\Delta,j,\bar{\jmath},R,r)$, where $(\Delta, j, \bar{\jmath})$ are the Dynkin labels of the conformal group and $(R,r)$ the Dynkin labels of the $R$-symmetry. The highest weight is, by definition, annihilated by the supercharges $\Sm$ and $\bar{\Sm}$ and the multiplet is then constructed by successive action of the Poincar\'e supercharges. Generic supermultiplets are called long multiplets and we will denote them by  $\mathcal{A}_{R,r\left( j,\bar{\jmath}\right)}^{\Delta}$ following the conventions of \cite{Dolan:2002zh}. Unitarity imposes restrictions on the conformal dimension of $\Am$ known as unitarity bounds. For generic long multiplets the bounds read,
\be 
\Delta \geq 2 + 2 j + 2 R + r\,, \, 2 + 2 \bar{\jmath} + 2R - r\, . 
\ee
If the highest weight is annihilated by some combination of the supercharges $\Qm$ and $\bar{\Qm}$ the multiplet shortens.
There are several types of shortening conditions depending on the Lorentz and $SU(2)_R$ quantum numbers of the charges that kill the highest weight, we denote them $\Bm$-type and $\Cm$-type shortening conditions.
\begin{eqnarray}
\Bm^{i} &:&\Qm_{\alpha }^{i} \Psi  =0, \\
\overline{\Bm}_{i} &:&\bar{\Qm}_{i\dot{\alpha}}\Psi =0, \\
\Cm^{i} &:&\left\{
\begin{array}{c}
\varepsilon ^{\alpha \beta }\Qm_{\alpha }^{i} \Psi_{\beta }=0,\quad j\neq 0 \\
\varepsilon ^{\alpha \beta }\Qm_{\alpha }^{i}\Qm_{\beta }^{i} \Psi =0,\quad j=0%
\end{array}%
\right. \\
\overline{\Cm}_{i} &:&\left\{
\begin{array}{c}
\varepsilon ^{\dot{\alpha}\dot{\beta}}\bar{\Qm}_{i\dot{\alpha}}
\Psi _{\dot{\beta}}=0,\quad \bar{\jmath}\neq 0 \\
\varepsilon ^{\dot{\alpha}\dot{\beta}}\bar{\Qm}_{i\dot{\alpha}}\bar{\Qm%
}_{i\dot{\beta}} \Psi =0,\quad \bar{\jmath}=0%
\end{array}%
\right.
\end{eqnarray}
$\Bm$-type conditions are sometimes called short while $\Cm$-type are sometimes called semi-short. In table \ref{shortening} we present all possible shortening conditions for the $\Nm=2$ superconformal algebra following the notation of \cite{Dolan:2002zh}.
\begin{table}[t]
\centering
\footnotesize
\begin{tabular}{|l|l l|l|}
\hline
Shortening & Quantum Number Relations&  & Multiplet 
\\ \hline\hline
$\mathcal{B}^{1}$ & $\Delta = 2R + r$ & $j=0$ & $\mathcal{B}_{R,r\left( 0,\bar{\jmath}%
\right) }$ \\ \hline
$\overline{\mathcal{B}}_{2}$ & $\Delta = 2R - r$ & $\bar{\jmath}=0$ & $\bar{\mathcal{B}}_{R,r\left( j,0\right)
} $ 
\\ \hline
$\mathcal{B}^{1}\cap \mathcal{B}^{2}$ & $\Delta = r$ & $R=0$  & $\mathcal{E}_{r\left( 0,%
\bar{\jmath}\right) }$ 
\\ \hline
$\overline{\mathcal{B}}_{1}\cap \overline{\mathcal{B}}_{2}$ & $\Delta = -r$ & $R=0$ & $\bar{\mathcal{E}}%
_{r\left( j,0\right) }$ 
\\ \hline
$\mathcal{B}^{1}\cap \overline{\mathcal{B}}_{2}$ & $\Delta = 2R$ & $j=\bar{\jmath}=r=0$ & $\hat{\mathcal{B}}_{R}$ 
\\
\hline\hline
$\mathcal{C}^{1}$ & $\Delta =2+2j+2R+r$  & & $\mathcal{C}_{R,r\left( j,%
\bar{\jmath}\right) }$ 
\\ \hline
$\overline{\mathcal{C}}_{2}$ & $\Delta =2+2\bar{\jmath}+2R-r$ &
& $\bar{\mathcal{C}}_{R,r\left( j,\bar{\jmath}\right) }$ 
\\ \hline
$\mathcal{C}^{1}\cap \mathcal{C}^{2}$ & $\Delta =2+2j+r$ & $R=0$ & $%
\mathcal{C}_{0,r\left( j,\bar{\jmath}\right) }$ 
\\ \hline
$\overline{\mathcal{C}}_{1}\cap \overline{\mathcal{C}}_{2}$ & $\Delta =2+2%
\bar{\jmath}-r$ & $R=0$ & $\bar{\mathcal{C}}_{0,r\left( j,\bar{\jmath}%
\right) }$ 
\\ \hline
$\mathcal{C}^{1}\cap \overline{\mathcal{C}}_{2}$ & $\Delta =2+2R+j+\bar{\jmath}$ & $r=\bar{\jmath}-j$ & $\hat{\mathcal{C}}_{R\left( j,\bar{\jmath}\right) }$
\\ \hline\hline
$\mathcal{B}^{1}\cap \overline{\mathcal{C}}_{2}$ & $\Delta =1+\bar{\jmath}%
+2R$ & $r=\bar{\jmath}+1$ & $\mathcal{D}_{R\left( 0,\bar{\jmath}\right) }$ 
\\ \hline
$\overline{\mathcal{B}}_{2}\cap \mathcal{C}^{1}$ & $\Delta =1+j+2R$ & 
$-r=j+1$ & $\bar{%
\mathcal{D}}_{R\left( j,0\right) }$ 
\\ \hline
$\mathcal{B}^{1}\cap \mathcal{B}^{2}\cap \overline{\mathcal{C}}_{2}$ & $%
\Delta =r=1+\bar{\jmath}$ & $R=0$ & $\mathcal{D}_{0\left( 0,%
\bar{\jmath}\right) }$ 
\\ \hline
$\mathcal{C}^{1}\cap \overline{\mathcal{B}}_{1}\cap \overline{\mathcal{B}}%
_{2}$ & $\Delta =-r=1+j$ & $R=0$ & $\bar{\mathcal{D}}_{0\left(
j,0\right) }$ 
\\ \hline
\end{tabular}
\caption{\label{shortening}Shortening conditions for the unitary irreducible representations of the $\Nm=2$ superconformal algebra.
} 
\end{table}
Among the most important short multiplets are the so-called chiral multiplets $\Em_{r}$ which obey two $\Bm$-type shortening conditions and are associated with the physics of the Coulomb branch of $\Nm=2$ theories. Also prominent are $1/2$ BPS multiplets, denoted by $\hat{\Bm}_R$, which obey two $\Bm$-type shortening conditions but of different chirality, these multiplets are associated with Higgs branch physics. 

We will be mostly interested in the multiplet $\hat{\Cm}_{0(0,0)}$. Multiplets of the type $\hat{\Cm}_{R(j,j)}$ obey semi-shortening conditions and the anti-commutation relation of the supercharges combine to give a generalized conservation equation. The special case $\hat{\Cm}_{0(0,0)}$ contains a spin two conserved current and we therefore identify it as the stress-tensor multiplet. In this work we will not consider theories that can be factorized as the product of two local theories, we will therefore assume a unique $\hat{\Cm}_{0(0,0)}$ multiplet. The multiplet also contains a conserved spin one operator which corresponds to the $SU(2)_R \times U(1)_r$ $R$-symmetry current.

Our goal is to study the super OPE of $\hat{\Cm}_{0(0,0)} \times \hat{\Cm}_{0(0,0)}$. In order to accomplish this we will carry out a detailed superspace analysis of three-point functions. Using $\Nm=2$ superspace language the stress-tensor multiplet 
can be represented by a superfield $\Jm$ that satisfies the conservation equation,
\begin{equation}
D^{\alpha i}D_{\alpha }^{j}\mathcal{J}=0\, ,\qquad \bar{D}_{%
\dot{\alpha}}^{i}\bar{D}^{j\dot{\alpha}}\Jm=0\, ,
\end{equation}%
where $D_{\a}^{i}$ and $\bar{D}_{\ad\, i}$ are $\Nm=2$ covariant derivatives and,
\be 
\Jm(x,\th,\thb) = J(x) + J_{\a \ad\, j}^{i}\th^{\a}_{i} \thb^{j\, \ad} + \ldots \, .
\ee
Both the scalar $J(x)$ and current $J^{(ij)}_{\a \ad}(x)$ will be of particular importance to us.

\paragraph{$2d$ chiral algebra and analytic bound on $c$.}

As we will see below, the super OPE expansion of $\hat{\Cm}_{0(0,0)}$ will allow us to obtain an analytic bound on the central charge $c$. Of prime importance in this analysis will be the existence of a protected subsector of observables present in any $\Nm=2$ SCFT, whose correlation functions are described by a $2d$ chiral algebra. We will review this construction with some detail in section 3, for now let us just give a short outline of the calculation. Four-dimensional operators described by the chiral algebra sit in multiplets of the type,
\be 
\label{Schurmultiplets}
\hat{\Bm}_{R}\, , \qquad \Dm_{R(0,\bar{\jmath})}\, , \qquad \bar{\Dm}_{R(j,0)}\, , \qquad \hat{\Cm}_{0(j,\bar{\jmath})}\, .
\ee
The $2d$ operator associated with the $\hat{\Cm}_{0(0,0)}$ multiplet is the $2d$ holomorphic stress-tensor, and it can be built using the $SU(2)_R$ current $J^{(ij)}_{\a \ad}(x)$,
\be 
J^{(ij)}_{\a \ad}(x) \to T(z)\, .
\ee
The $2d$ stress-tensor correlator constitutes a solvable truncation of the full four-point function of four currents  $J^{(ij)}_{\a \ad}(x)$, and can be completely fixed by symmetry. This correlator can be expanded in conformal blocks associated with the multiplets listed in \eqref{Schurmultiplets}, and unitarity of the four-dimensional theory implies an analytic bound on $c$ valid for any interacting $\Nm=2$ SCFT.

\paragraph{Crossing symmetry and numerical bounds on $a/c$.}

The supersymmetric selection rules are also relevant for the crossing symmetry equation of the superconformal primary $J(x)$. This is a more challenging calculation and we will only present some partial results. The motivation for this is that using numerical bootstrap techniques we would then have access to the $a$-anomaly coefficient. This coefficient plays no role in the $2d$ chiral algebra, and cannot be bounded analytically, at least not with the techniques used in this paper. To obtain bounds on $a$ one has to resort to numerics. In order to write crossing symmetry a fundamental ingredient is the conformal block expansion of the $\langle J(x_1)J(x_2)J(x_3)J(x_4)\rangle$ correlator. An attractive challenge for the $\Nm=2$ bootstrap program would be to recover or even improve on the bounds found in \cite{Hofman:2008ar} whose supersymmetric version reads,
\be 
\frac{1}{2} \leq \frac{a}{c} \leq \frac{5}{4}\, .
\ee
In section 5 the super OPE selection rules will help us understand how the different $\Nm=2$ multiplets contribute to the $J$ correlator, a necessary first step before a crossing symmetry equation can be written. 

\section{Three-point functions}

We will now study all possible three point functions  $\langle \Jm \Jm \Om \rangle$ between two stress-tensor multiplets and a third arbitrary operator. The correlator for three stress-tensor multiplets $\langle \Jm \Jm \Jm \rangle$ was studied by Kuzenko and Theisen in \cite{Kuzenko:1999pi}. We will use their notation and borrow some of their results. The starting point is the general expression for three-point functions in $\Nm=2$ superspace \cite{Osborn:1998qu,Park:1999pd}
\be 
\langle \Jm(z_1) \Jm(z_2) \Om^{\Im}(z_3) \rangle = \frac{1}{(x_{\bar{1}3})^2 (x_{\bar{3}1})^2 (x_{\bar{2}3})^2 (x_{\bar{3}2})^2} H^{\Im}(\mathbf{Z}_3)\, \label{JJO},
\ee
where $\Im = (\alpha, \dot{\alpha},R,r)$ is a collective index that labels the irreducible representation to which $\Om$ belongs. The (anti-)chiral combinations of coordinates are,
\begin{align}
& x^{\ad \a}_{\bar{1}2}  = - x^{\ad \a}_{2\bar{1}} = x^{\ad \a}_{1-} -  x^{\ad \a}_{2+} -4\mathrm{i}\, \theta^{\a}_{2\,i} \bar{\theta}^{\ad i}_1\, ,
\\
& \theta_{12} = \theta_1 - \theta_2\, , \qquad  \bar{\theta}_{12} = \bar{\theta}_1 - \bar{\theta}_2\, ,
\end{align}
with $x^{\ad \a}_{\pm} = x^{\ad \a} \mp 2{\rm i}\th_{i}^\a \thb^{\ad\, i}$. The argument of $H$ is given by three superconformally covariant coordinates $\mathbf{Z}_3 = (\Xbf_3, \Theta_3, \bar{\Theta}_3)$,
\begin{align}
&
\mathbf{X}_{3\,\a\, \ad}  = \frac{x_{3\bar{1}\,\a \bd}{ x_{\bar{1}2}^{\bd\b} }x_{2\bar{3}\,\b\ad}}{(x_{3\bar{1}})^2 (x_{2\bar{3}})^2}\, ,
&
&
\bar{\mathbf{X}}_{3\,\a\ad} = \mathbf{X}^{\dagger}_{3\,\a\ad}  
= -\frac{x_{3\bar{2}\, \a\bd} x_{\bar{2}1}^{\bd\b} x_{1\bar{3}\,\b\ad}}{(x_{3\bar{2}})^2  (x_{1\bar{3}})^2}\, ,
&
\\
&
\Theta^{i}_{3\, \a} = \mathrm{i} \left(\frac{x_{\bar{2}3\, \a \ad}}{x_{\bar{2}3}^2}\bar{\theta}^{\,\ad i}_{32} 
- \frac{x_{\bar{1}3\, \a \ad}}{x_{\bar{1}3}^2}\bar{\theta}^{\,\ad i}_{31} \right)\, ,
&
&
\bar{\Theta}_{3\, \ad\, i} = \mathrm{i} \left(\theta^{\a}_{32\, i} \frac{x_{\bar{3}2\, \a \ad}}{x_{\bar{3}2}^2} 
- \theta^{\a}_{31\, i} \frac{x_{\bar{3}1\, \a \ad}}{x_{\bar{3}1}^2} \right)\, \label{XXBTTb}.
&
\end{align}
An important relation which will play a key role in our computations is
\be
\bar{\mathbf{X}}_{3\,\a\ad} = \mathbf{X}_{3\,\a\ad}-4 \mathrm{i}\, \Theta^{i}_{3\,\a} \bar{\Theta}_{3\,\ad\, i}\, .
\ee
In addition, the function $H$ satisfies the scaling condition,
\be 
\label{scaling}
H^{\Im}(\lambda \bar{\lambda}\Xbf_3,\lambda \Theta_3,\bar{\lambda} \bar{\Theta}_3) = \lambda^{2a} \bar{\lambda}^{2\bar{a}} H^{\Im}(\Xbf_3,\Theta_3,\bar{\Theta}_3)\, , 
\ee
with $a-2\bar{a} = 2-q$ and $\bar{a}-2a = 2-\bar{q}$, where $\Delta=q+\bar{q}$ and $r=q-\bar{q}$. Extra restrictions are obtained by imposing the conservation equations of $\Jm$, these imply,
\begin{align}
\label{dTheta2}
&
\frac{\pd^2}{\pd \Theta^{i}_{3\,\a} \pd \Theta^{\a\, j}_{3}} H^{\Im}(\mathbf{Z}_3) = 0\, ,
&
&
\frac{\pd^2}{\pd \bar{\Theta}^{\ad}_{3\,i} \pd \bar{\Theta}}_{3\,\ad\, j} H^{\Im}(\mathbf{Z}_3) = 0\, ,
&
\\
\label{D2}
&
\quad \, \, \, \, \Dm^{\a}_{i} \Dm_{\a\, j}\, H^{\Im}(\mathbf{Z}_3) = 0 \, ,
&
&
\quad \, \, \, \, \, \, \tilde{\Dm}^{\ad\, i} \tilde{\Dm}_{\ad}^{j}\, H^{\Im}(\mathbf{Z}_3) = 0\, ,
&
\end{align}
where
\be
\label{Ds}
\Dm_{\ad\, i} =  \frac{\pd}{\pd \Theta^{\a\, i}_3} +4\mathrm{i} \bar{\Theta}^{\ad}_{3\,i} \frac{\pd}{\pd \Xbf_3^{\ad \a}}\, ,
\qquad
\tilde{\Dm}^{\ad\, i} =  \frac{\pd}{\pd \bar{\Theta}_{3\, \ad\, i}} -4\mathrm{i} \Theta^{i}_{3\,\a} \frac{\pd}{\pd \Xbf_{3\, \a \ad}}\, .
\ee
In order to see how these restrictions are imposed, let us work out an example in detail. For an operator $\Om$ which is a scalar under Lorentz and $SU(2)_R \times U(1)$, equations \eqref{dTheta2} imply that $H$ can be at most quadratic in $\Theta_3$ and $\bar{\Theta}_3$. Thus, we will consider the following ansatz:
\be 
H(\mathbf{Z_3}) = f(\mathbf{X}_3) + g_{\alpha \dot{\alpha}}(\mathbf{X}_3) \Theta^{\alpha}_3\bar{\Theta}^{\dot{\alpha}}_3  + h_{\alpha \beta \dot{\alpha} \dot{\beta}}(\mathbf{X}_3) \Theta_{3}^{\alpha \beta}\bar{\Theta}_{3}^{\dot{\alpha} \dot{\beta}}\, \label{ansatz1},
\ee
where
\be 
\Theta_{3}^{\a \b} = \Theta_{3}^{\a i}\Theta_{3}^{\b j}\,\veps_{i j}\, , \qquad 
\bar{\Theta}^{\ad \bd}_{3} = \bar{\Theta}^{\ad}_{3\,i}\Theta^{\bd}_{3\,j}\,\veps^{i j}\, .
\ee
This is the most general expression consistent with $SU(2)_R \times U(1)_r$ invariance quadratic in the $\Theta$s. Next, we impose the scaling condition \eqref{scaling},
\be 
H(\lambda \bar{\lambda}\mathbf{X}_3,\lambda  \Theta_3, \bar{\lambda}\bar{\Theta}_3) = \lambda^{-4-\Delta/2} \bar{\lambda}^{-4-\Delta/2}H(\mathbf{X}_3, \Theta_3, \bar{\Theta}_3)\, .
\ee
Hence, the functions $f$, $g$, and $h$ are known up to an overall constant:
\be 
\label{ansatzlong}
H(\mathbf{Z_3}) = a_1 \frac{1}{(\Xbf_3^2)^{2-\frac{\Delta}{2}}} + a_2 \frac{\Theta_3^{\alpha} \Xbf_{3\,\alpha \dot{\alpha}}\bar{\Theta}_3^{\dot{\alpha}}}{(\Xbf_3^2)^{3-\frac{\Delta}{2}}} +  a_3 \frac{\Theta^{\alpha \beta} \Xbf_{3\,\alpha \dot{\alpha}}\Xbf_{3\,\beta \dot{\beta}}\bar{\Theta}^{\dot{\alpha} \dot{\beta}}}{(\Xbf_3^2)^{4-\frac{\Delta}{2}}} \, .
\ee
Our correlator \eqref{JJO} should also be invariant under the exchange $z_1 \leftrightarrow z_2$ which implies $(\Xbf_3, \Theta_3, \bar{\Theta}_3)\to (-\bar{\Xbf}_3,-\Theta_3,-\bar{\Theta}_3)$, as can be checked from \eqref{XXBTTb}. We will call this symmetry $\mathbb{Z}_2$ for short. Then,
\be 
\label{Z2condition}
H(\Xbf_3, \Theta_3, \bar{\Theta}_3) = H(-\bar{\Xbf}_3,-\Theta_3,-\bar{\Theta}_3)\, .
\ee
This condition turns out to be very restrictive. In particular, if a function satisfies \eqref{dTheta2} and the $\mathbb{Z}_2$ condition, it also satisfies equations \eqref{D2}. Fixing the correlator is now a standard exercise in Grassmann algebra, we Taylor expand \eqref{Z2condition} in powers of the Grassmann variables and equate coefficients in both sides in order to fix $(a_1,a_2,a_3)$. Details of our calculations along with some superspace identities are presented in appendix B.

For arbitrary $\Delta$ there is a unique solution given by
\be 
(a_1,a_2,a_3) = c_{\Jm \Jm \Om}(1\, , \,\mathrm{i}(\Delta-4)\, , -\tfrac{1}{3}(\Delta-4) (\Delta-6))\, .\label{result1}
\ee
This solution is our generalization of the $\langle \Jm \Jm \Jm \rangle$ correlator for the case in which the third operator is a long multiplet $\Am^{\Delta}_{0,0(0,0)}$ with unrestricted conformal dimension $\Delta$.

For the special case $\Delta=2$ the long multiplet hits its unitarity bound and splits according to,
\be 
\label{multipeltsplit}
\Am^{2}_{0,0(0,0)} = \hat{\Cm}_{0(0,0)} +\Dm_{1(0,0)} +\bar{\Dm}_{1(0,0)}+\hat{\Bm}_2\, .
\ee
The results of this section imply that $\Dm$ and $\hat{\Bm}$ multiplets are not allowed.\footnote{We refer the reader to \eqref{superOPE} where we have collected in a single equation the super OPE selections rules obtained in this section.} Then, for $\Delta=2$ the only surviving term in \eqref{multipeltsplit} is $\hat{\Cm}_{0(0,0)}$, and we just recover the $\langle \Jm \Jm \Jm \rangle$ correlator solution:
\be 
(a_1,a_2,a_3) = c^{(1)}_{\Jm \Jm \Jm}(1\,, -2\, \mathrm{i}\,, 0) + c^{(2)}_{\Jm \Jm \Jm}(0\,, 0\,, 1)\, .
\ee
That is, there are two independent structures. These two structures can be associated to the $a$ and $c$ anomaly coefficients, the exact relations were worked out in \cite{Kuzenko:1999pi},
\be 
c^{(1)}_{\Jm \Jm \Jm} = \frac{3}{32 \pi^6}(4\, a - c)\, , \qquad c^{(2)}_{\Jm \Jm \Jm} = \frac{1}{8 \pi^6}(4\, a - 5\, c)\, .
\ee
The presence of two parameters is due to the fact that the last term in \eqref{ansatzlong} is automatically symmetric under $z_1 \leftrightarrow z_2$ when $\Delta=2$,\footnote{From now on we will ignore the subindex 3 in $(\Xbf_3, \Theta_3, \bar{\Theta}_3)$.}
\be 
\frac{\Theta^{\alpha \beta} \Xbf_{\alpha \dot{\alpha}}\Xbf_{\beta \dot{\beta}}\bar{\Theta}^{\dot{\alpha} \dot{\beta}}}{(\Xbf^2)^{3}}  = \frac{\Theta^{\alpha \beta} \bar{\Xbf}_{\alpha \dot{\alpha}}\bar{\Xbf}_{\beta \dot{\beta}}\bar{\Theta}^{\dot{\alpha} \dot{\beta}}}{(\bar{\Xbf}^2)^{3}} \, .
\ee
Another way to phrase this, is that there is a ``nilpotent invariant'', namely, a purely fermionic term that satisfies all the symmetry requirements. It implies that we can not reconstruct the full superspace three-point function starting from the three-point function of the superconformal primaries \cite{Osborn:1998qu}. This is a generic property of superconformal field theories, unlike the pure conformal case in which three-point functions of descendants can always be obtained from that of primaries by taking derivatives. Although nilpotent invariants are to be expected, for some special cases it is impossible to build three-point invariants that satisfy all the symmetries of the correlator. Well known cases are $1/2$ BPS operators in $\Nm=2$ and $\Nm=4$ theories \cite{Arutyunov:2001qw,Dolan:2001tt,Nirschl:2004pa,Dolan:2004mu} and (anti)chiral operators in $\Nm=2$ and $\Nm=1$ theories \cite{Poland:2010wg,Fitzpatrick:2014oza}. As we will see below, nilpotent invariants will also be present when we consider operators with spin.

\noindent
\textbf{$\Nm=1$ check}: As a check on our result, let us reduce it to $\Nm=1$ superspace language and compared it the known solutions of \cite{Khandker:2014mpa,Berkooz:2014yda,Fortin:2011nq}. Using the coefficients \eqref{result1} in \eqref{ansatz1} and rewriting in denominators in terms of $\Xbf \cdot \bar{\Xbf}$. Setting the $i=2$ components to zero we obtain,
\be
\Theta_{\a\, i=1} \to \Theta_{\a}\, , \qquad  \Theta_{\a\, i=2} \to 0\, ,
\ee
where $\Theta_{\a}$ is the analogous $\Nm=1$ coordinate. Our solution reduces to,
\be
H(\mathbf{Z}) = \frac{1}{(\Xbf \cdot \bar{\Xbf})^{2-\frac{\Delta}{2}}} \left(1 -\frac{1}{4}(\Delta-4)(\Delta-6) \frac{\Theta^2 \bar{\Theta}^2}{(\bar{\Xbf} \cdot \Xbf)} \right)\, ,
\ee
in perfect agreement with the $\Nm=1$ result of \cite{Fortin:2011nq,Berkooz:2014yda,Khandker:2014mpa}.

The procedure is now clear:
\begin{itemize}
\item Write the most general ansatz consistent with \eqref{dTheta2}.
\item Fix the $\Xbf$-dependence using the scaling condition \eqref{scaling}
\item Fix the arbitrary coefficients by imposing the $\mathbb{Z}_2$ symmetry \eqref{Z2condition}.
\end{itemize}
We now apply this strategy to all possible combinations of Lorentz and $SU(2)_R \times U(1)_r$ quantum numbers in order to find the $\Nm=2$ selection rules for the OPE of two stress-tensor multiplets.

\subsection{Solutions}

\subsubsection*{$\Am^{\Delta}_{0,0(\frac{\ell}{2},\frac{\ell}{2})}$}
The most general ansatz for arbitrary $\ell$ consistent with the conditions discussed above is
\begin{align}
\label{ansatzlongell1}
\begin{split}
H(\mathbf{Z}) & = \frac{ \Xbf_{\a_1 \ad_1} \ldots \Xbf_{\a_{\ell} \ad_{\ell}}}{(\Xbf^2)^{2-\frac{\Delta-\ell}{2}}}\left( a_1 + a_2\, \frac{\Theta^{\alpha i}\Xbf_{\alpha \dot{\alpha}}\bar{\Theta}^{\dot{\alpha}}_i}{\Xbf^2} 
+  a_3\, \frac{\Theta^{\alpha \beta}\Xbf_{\alpha \dot{\alpha}}\Xbf_{\beta \dot{\beta}}\bar{\Theta}^{\dot{\alpha} \dot{\beta}}}{(\Xbf^2)^2}\right)
\\
& + \frac{\Xbf_{\a_2 \ad_2} \ldots \Xbf_{\a_{\ell} \ad_{\ell}}}{(\Xbf^2)^{2-\frac{\Delta-\ell}{2}}}\left(a_4\, \Theta^{i}_{\a_1}\bar{\Theta}_{\ad_1\, i} + a_5\, \frac{\bar{\Theta}_{\ad_1 \bd}\Xbf^{\bd \b}\Theta_{\b \a_1}}{\Xbf^2}\right)\,\\
& +a_6 \frac{\Xbf_{\a_3 \ad_3} \ldots \Xbf_{\a_{\ell} \ad_{\ell}}}{(\Xbf^2)^{2-\frac{\Delta-\ell}{2}}}\Theta_{\a_1\a_2}{\bar\Theta}_{\ad_1\ad_2}\, ,
\end{split}
\end{align}
where it is understood that the indices $(\a_1, \ldots, \a_{\ell})$ and $(\ad_1, \ldots, \ad_{\ell})$ are symmetrized with weight one. Imposing the $\mathbb{Z}_2$ symmetry we find, for the odd case,
\begin{align}
\vec{a} = & c_{\Jm \Jm \Om}\left(0,\frac{1}{2(\Delta-\ell)},\frac{\mathrm{i}(\Delta-6-\ell)}{4(\Delta-2)},\frac{1}{\Delta-4-\ell},\frac{\mathrm{i}(\Delta-2-\ell)}{2(\Delta-2)},\frac{\mathrm{i}(1-\ell)}{(\Delta-4-\ell)} \right).
\end{align}
For $\ell=1$ the last structure in \eqref{ansatzlongell1} can not contribute.

For the $\ell$ even case we find two different solutions
\begin{align}
\begin{split}
\vec{a} & =  c^{(1)}_{\Jm \Jm \Om}\left(0,0,\frac{1}{2}(\Delta-6-\ell),(3\Delta+\ell-6),0,\frac{(3(\Delta-2)^2-2\ell-\ell^2)}{(\Delta-4-\ell)} \right)
\\
& + c^{(2)}_{\Jm \Jm \Om}\left(\mathrm{i}\frac{1}{2\ell},-\frac{(\Delta-4-\ell)}{2\ell},-\mathrm{i}\frac{(\Delta+\ell-2)(\Delta-4-\ell)(\Delta-6-\ell)}{2\ell(3\Delta+\ell-6)},\right.\\ &\phantom{+}\left.1,0,-2\mathrm{i}\frac{(\Delta-3)(\Delta-2+\ell)}{(3\Delta+\ell-6)} \right)
\end{split}
\end{align}
The two-parameter solution is due to the existence of three-point ``nilpotent invariant'' that can only be constructed when the spin is even. Indeed, the object
\begin{align}
&\frac{1}{2}(\Delta-6-\ell)\frac{ \Xbf_{\a_1 \ad_1} \ldots \Xbf_{\a_{\ell} \ad_{\ell}}}{(\Xbf^2)^{4-\frac{\Delta-\ell}{2}}}\, \Theta^{\alpha \beta}\Xbf_{\alpha \dot{\alpha}}\Xbf_{\beta \dot{\beta}}\bar{\Theta}^{\dot{\alpha} \dot{\beta}}
+(3\Delta+\ell-6)\frac{\Xbf_{\a_2 \ad_2} \ldots \Xbf_{\a_{\ell} \ad_{\ell}}}{(\Xbf^2)^{2-\frac{\Delta-\ell}{2}}}\, \Theta^{i}_{\a_1}\bar{\Theta}_{\ad_1\, i} \nonumber\\
&+\frac{(3(\Delta-2)^2-2\ell-\ell^2)}{(\Delta-4-\ell)} \frac{\Xbf_{\a_3 \ad_3} \ldots \Xbf_{\a_{\ell} \ad_{\ell}}}{(\Xbf^2)^{2-\frac{\Delta-\ell}{2}}}\Theta_{\a_1\a_2}{\bar\Theta}_{\ad_1\ad_2}\, ,
\end{align}
satisfies all the constraints imposed by $\Nm=2$ superconformal symmetry. As a consequence, the superconformal block will have an undetermined parameter. In \cite{Khandker:2014mpa} superconformal blocks for general scalar operators were obtained where the same happens, the block has a number of free parameters, in that case things can be improved if one imposes conservation or chirality conditions. Our result implies that in $\Nm=2$ theories, even imposing the conservation condition is not enough, and there will be an unfixed parameter in the superconformal block expression. It would be interesting to understand whether this parameter has some physical meaning, like in the $\Delta=2$, $\ell=0$ case, where they are identified with anomaly coefficients.

At the unitarity bound we have the splitting,
\be 
\Am^{2+\ell}_{0,0(\frac{\ell}{2},\frac{\ell}{2})} = \hat{\Cm}_{0(\frac{\ell}{2},\frac{\ell}{2})}
+\hat{\Cm}_{\frac{1}{2}(\frac{\ell-1}{2},\frac{\ell}{2})}
+\hat{\Cm}_{\frac{1}{2}(\frac{\ell}{2},\frac{\ell-1}{2})}
+\hat{\Cm}_{1(\frac{\ell-1}{2},\frac{\ell-1}{2})}\, .
\ee
The multiplets $\hat{\Cm}_{\frac{1}{2}(\frac{\ell-1}{2},\frac{\ell}{2})}$ and $\hat{\Cm}_{\frac{1}{2}(\frac{\ell}{2},\frac{\ell-1}{2})}$ are not allowed by the selection rules (see \eqref{superOPE}). The $\hat{\Cm}_{0(\frac{\ell}{2},\frac{\ell}{2})}$ multiplets contain higher spin currents and are not expected to appear in interacting theories, with the exception of $\ell=0$.

\subsubsection*{$\Am^{\Delta}_{0,0(\frac{\ell+2}{2},\frac{\ell}{2})}$}

We also found solutions for complex long multiplets,
\begin{align}
\begin{split}
H(\mathbf{Z}) & = \frac{ \Xbf_{\a_1 \ad_1} \ldots \Xbf_{\a_{\ell} \ad_{\ell}}}{(\Xbf^2)^{3-\frac{\Delta-\ell}{2}}}\left( a_1\, \Theta^{i}_{\a_{\ell+1}}\Xbf_{\a_{\ell+2}\, \ad}\bar{\Theta}^{\ad}_{\, i} + a_2\,\epsilon_{\a_{\ell+1}\, \a}\Xbf_{\a_{\ell+2}\, \ad} \frac{{\Theta}^{\a \b}\Xbf_{\b \bd}{\bar\Theta}^{\bd \ad}}{\Xbf^2}\right)\,\\
& +a_3 \frac{\Xbf_{\a_1 \ad_1} \ldots \Xbf_{\a_{\ell-1}\, \ad_{\ell-1}}}{(\Xbf^2)^{3-\frac{\Delta-\ell}{2}}}\Theta_{\a_{\ell}\,\a_{\ell+1}}\Xbf_{\a_{\ell+2}\,\ad}\epsilon^{\ad\bd}{\bar\Theta}_{\ad_{\ell}\,\bd}\, .
\end{split}
\end{align}
For $\ell$ even we have $(a_1,a_2,a_3)=c_{\Jm \Jm \Om}\left(0,(\Delta-6-\ell),2(\Delta-2)\right)$, while for $\ell$ odd $(a_1,a_2,a_3)=c_{\Jm \Jm \Om}\left(2,\mathrm{i}(\Delta-6-\ell),-2\mathrm{i}\ell\right)$. For $\ell=0$ there is no solution. 

At the unitarity bound we have,
\be 
\Am^{3+\ell}_{0,0(\frac{\ell+2}{2},\frac{\ell}{2})} = \Cm_{0,0(\frac{\ell+2}{2},\frac{\ell}{2})}
+\Cm_{\frac{1}{2},\frac{1}{2}(\frac{\ell+1}{2},\frac{\ell}{2})}\, .
\ee
The multiplet $\Cm_{\frac{1}{2},\frac{1}{2}(\frac{\ell+1}{2},\frac{\ell}{2})}$ is not allowed by the selection rules (see \eqref{superOPE}).

\subsubsection*{$\Am^{\Delta}_{0,0(\frac{\ell+4}{2},\frac{\ell}{2})}$}

Finally, there is another long multiplet
\be
H(\mathbf{Z}) = \frac{c_{\Jm \Jm \Om}}{(\Xbf^2)^{4-\frac{\Delta-\ell}{2}}}
\Xbf_{\a_1\ad_1} \ldots \Xbf_{\a_{\ell} \ad_{\ell}} \Xbf_{\a_{\ell+1} \ad } \Xbf_{\a_{\ell+2} \bd} 
 \Theta_{\a_{\ell+3} \a_{\ell+4}}\bar{\Theta}^{\ad \bd}\, ,
\ee
with $c_{\Jm \Jm \Om} \neq 0$ only for $\ell$ even. 

At the unitarity bound we have,
\be 
\Am^{4+\ell}_{0,0(\frac{\ell+4}{2},\frac{\ell}{2})} = \Cm_{0,0(\frac{\ell+4}{2},\frac{\ell}{2})}
+\Cm_{\frac{1}{2},\frac{1}{2}(\frac{\ell+3}{2},\frac{\ell}{2})}\, .
\ee
The multiplet $\Cm_{\frac{1}{2},\frac{1}{2}(\frac{\ell+3}{2},\frac{\ell}{2})}$ is not allowed by the selection rules (see \eqref{superOPE}).

\subsubsection*{$\bar{\Cm}_{0,-3(\frac{\ell+2}{2},\frac{\ell}{2})}$}

We also found solutions that fix the conformal dimension $\Delta$,
\be
H(\mathbf{Z})=\frac{c_{\Jm \Jm \Om}}{\left(\Xbf^2\right)^{\frac{3}{2}-\frac{\Delta-\ell}{2}}}\Xbf_{\a_1\ad_1}\cdots \Xbf_{\a_\ell \ad_\ell} 
\Theta_{\a_{\ell+1}\a_{\ell+2}}\, ,
\ee
has nonzero $a$ for $\Delta=5+\ell$, and $\ell\ge 0$ even. This is precisely the unitarity bound for this quantum numbers and corresponds to a semi-short multiplet of the $\Cm$-type.

\subsubsection*{$\bar{\Cm}_{\frac{1}{2},-\frac{3}{2}(\frac{\ell+1}{2},\frac{\ell}{2})}$}

We can also have multiplets that transform non-trivially under $SU(2)_R$ representations,
\begin{align}
H(\mathbf{Z}) & =\frac{ \Xbf_{\a_1\ad_1} \cdots \Xbf_{\a_{\ell-1}\ad_{\ell-1}}}{\left(\Xbf^2\right)^{\frac{9}{4}-\frac{\Delta-\ell}{2}}}
\left(
a_1 \Xbf_{\a_{\ell} \ad_{\ell}} \Theta^{i}_{\a_{\ell+1}} 
+ a_2 \frac{\Xbf_{\a_{\ell} \ad_{\ell}} \Xbf^{\bd \b}}{\Xbf^2} \Theta_{\a_{\ell+1} \b}\bar{\Theta}^{i}_{\bd}
+ a_3 \Theta_{\a_{\ell} \a_{\ell+1}}\bar{\Theta}^{i}_{\ad_{\ell}}
\right)\, .
\end{align}
$H$ is nonvanishing only for $\Delta=\frac{9}{2}+\ell$ which is the unitarity bound for these quantum numbers. The solution is $(a_1,a_2,a_3)=c_{\Jm \Jm \Om}\,(1,0,\mathrm{i}\, \ell )$ for $\ell$ odd and $(a_1,a_2,a_3)=c_{\Jm \Jm \Om}\,(0,1,0)$ for $\ell$ even.

\subsubsection*{$\hat{\Cm}_{1(\frac{\ell}{2},\frac{\ell}{2})}$}

For $SU(2)_R$ triplets we find the following family,
\be
H = \frac{c_{\Jm \Jm \Om}}{\left(\Xbf^2\right)^{2-\frac{\Delta-\ell}{2}}}\Xbf_{\a_1 \ad_1} \cdots \Xbf_{\a_{\ell-1}\ad_{\ell-1}}\Theta^{(i}_{\a_\ell}\bar{\Theta}^{j)}_{\ad_\ell}
\ee
This structure is nonvanishing only for $\Delta=4+\ell$, which is again the unitarity bound. Now, the flavor current sits in $\hat{\Bm}_{1}$ multiplet which is a triplet under $SU(2)_R$ and has $\ell=0$. Its superspace field was denoted by $L^{ij}$ in \cite{Kuzenko:1999pi} and it was found that $\langle \Jm \Jm L^{ij} \rangle = 0$. Our solution is consistent with their result.

For $SU(2)_R$ representations higher than $R=1$ no solutions exist due to the condition that the correlator be at most quadratic in $\Theta$ and $\bar{\Theta}$. 

\subsection{Extra solutions}

In addition to the multiplets described above we found extra solutions.

\subsubsection*{Non-unitary $\left(R,r,j,\bar{\jmath}\right)=\left(\frac{1}{2},-\frac{3}{2},\frac{\ell}{2},\frac{\ell+1}{2} \right)$ solution}

The following structure is also allowed,
\begin{align}
\begin{split}
H=&\frac{\Xbf_{\a_1\ad_1} \cdots \Xbf_{\a_{\ell-1}\ad_{\ell-1}}}{\left(\Xbf^2\right)^{\frac{11}{4}-\frac{\Delta-\ell}{2}}}
\Bigg(
a_1 \Xbf_{\a_\ell \ad_\ell} \Xbf_{\a \ad_{\ell + 1}} \Theta^{\a\, i}
+ a_2\frac{\Xbf_{\a_\ell \ad_\ell}\Xbf_{\a \ad_{\ell+1}}\Xbf_{\b \bd} \Theta^{\a \b} \bar{\Theta}^{\bd\, i}}{\Xbf^2}\Bigg. \\ &\Bigg. 
+ a_3 \Xbf_{\a \ad_\ell} \Theta^{\a}_{\ph{\a}\a_{\ell}} \bar{\Theta}^{i}_{\ad_{\ell+1}} 
\Bigg)
\end{split}
\end{align}
$H$ is nonvanishing only for $\Delta=\frac{3}{2}-\ell$ for $(a_1,a_2,a_3) = c_{\Jm \Jm \Om}(1,2\,\mathrm{i},0)$ for $\ell=0$, $(a_1,a_2,a_3) = c_{\Jm \Jm \Om}(0,\ell+2,\ell)$ for $\ell$ odd, and $(a_1,a_2,a_3) = c_{\Jm \Jm \Om}(1,\mathrm{i}\,(\ell+2),\mathrm{i}\, \ell)$ for $\ell$ even. This solution is below the unitarity bound and therefore of no interest to us. Similar non-unitary solutions were found in \cite{Berkooz:2014yda}.

\subsubsection*{Non-unitary $\left(R,r,j,\bar{\jmath}\right)=\left(\frac{1}{2},-\frac{3}{2},\frac{\ell}{2},\frac{\ell+3}{2} \right)$ solution}

We also found
\be
H(\mathbf{Z}) = c_{\Jm \Jm \Om} \frac{\Xbf_{\a_1\ad_1}\cdots \Xbf_{\a_{\ell}\ad_{\ell}}}{\left(\Xbf^2\right)^{\frac{19}{4}-\frac{\Delta-\ell}{2}}}\Xbf_{\a \ad_{\ell+1}}X_{\b \ad_{\ell+2}} \Theta^{\a \b} \bar{\Theta}^{i}_{\ad_{\ell+3}}\, .
\ee
$H$ is nonvanishing only for $\Delta=\frac{3}{2}-\ell$ and only for $\ell\ge 1$ odd. As the case above, this is below the unitarity bound and has no relevance for this work.

\subsubsection*{$\Am^{\frac{13}{2}+\ell}_{\frac{1}{2},-\frac{3}{2}(\frac{\ell+3}{2},\frac{\ell}{2})}$}

Finally, we found a strange solution that corresponds to a long multiplet with fixed conformal dimension:
\be
H(\mathbf{Z}) = \frac{c_{\Jm \Jm \Om}}{\left(\Xbf^2\right)^{\frac{13}{4}-\frac{\Delta-\ell}{2}}}\Xbf_{\a_1\ad_1}\cdots \Xbf_{\a_{\ell}\ad_{\ell}}\Xbf_{\a_{\ell+1} \ad} \Theta_{\a_{\ell+2} \a_{\ell+3}} \bar{\Theta}^{\ad\, i}\, .
\ee
The only restriction for this long multiplet is that the conformal dimension be above the $\Nm=2$ unitarity bound $\Delta = \frac{9}{2}+\ell$, it is then puzzling that our solution fixes its dimension to $\Delta=\frac{13}{2}+\ell$. Because it sits above the unitarity bound we can not interpret it as a contribution from a short multiplet. One possible explanation is that this multiplet corresponds to a theory that has enhanced $\Nm=4$ symmetry. $\Nm=2$ long multiplets with fixed conformal dimension appear if one decomposes $\Nm=4$ multiplets. The OPE of two $\Nm=4$ stress-tensors is well known \cite{Dolan:2001tt,Nirschl:2004pa},
\begin{align}
\begin{split}
\label{N4OPE}
\Bm_{\lbrack0,2,0\rbrack} \times \Bm_{\lbrack0,2,0\rbrack} & \sim \Bm_{\lbrack0,2,0\rbrack}+\Bm_{\lbrack0,4,0\rbrack}+\Bm_{\lbrack1,0,1\rbrack}+\Bm_{\lbrack1,2,1\rbrack}+\Bm_{\lbrack2,0,2\rbrack}
\\
&
+\Cm_{\lbrack0,0,0
\rbrack, \ell}+\Cm_{\lbrack1,0,1
\rbrack,\ell}+\Cm_{\lbrack0,2,0
\rbrack,\ell}+\ldots\, ,
\end{split}
\end{align}
where the $\ldots$ stand for long multiplets with unrestricted conformal dimension. In the decomposition of the $\Nm=4$ stress-tensor multiplet we find, among other things, the $\Nm=2$ stress-tensor multiplet,
\be 
\Bm_{\lbrack0,2,0
\rbrack} = \ldots + \hat{\Cm}_{0(0,0)} + \ldots \, .
\ee
Our curious multiplet could appear in the decomposition of one of the multiplets in the RHS of \eqref{N4OPE}. The $\Bm_{\lbrack0,p,0
\rbrack}$ decompositions were worked out in \cite{Dolan:2002zh} and our multiplet does not appear there, our guess is that is hiding somewhere in the $\Cm$ multiplets. In principle one could use the character techniques of \cite{Bianchi:2006ti} to confirm this suspicion, although straightforward, this type of calculation can still become quite involved. In the remainder, we will ignore this solution considering it an accident with no relevance to $\Nm=2$ dynamics.

\section{2d chiral algebra and central charge bound}

The superspace analysis of the previous section allows us to write the super OPE selection rules for the $\Nm=2$ stress-tensor multiplet,\footnote{To avoid cluttering we do not write the conjugate multiplets.}
\begin{align}
\label{superOPE}
\begin{split}
\hat{\Cm}_{0(0,0)} \times \hat{\Cm}_{0(0,0)} & \sim \Im+ \hat{\Cm}_{0(\frac{\ell}{2},\frac{\ell}{2})}
 + \hat{\Cm}_{1(\frac{\ell}{2},\frac{\ell}{2})}  +\Cm_{\frac{1}{2},\frac{3}{2}(\frac{\ell}{2},\frac{\ell+1}{2})}
\\
&
+\Cm_{0,3(\frac{\ell}{2},\frac{\ell+2}{2})}
+\Cm_{0,0(\frac{\ell+2}{2},\frac{\ell}{2})}
+\Cm_{0,0(\frac{\ell+4}{2},\frac{\ell}{2})}
\\
&
+\Am^{\Delta}_{0,0(\frac{\ell}{2},\frac{\ell}{2})}
+\Am^{\Delta}_{0,0(\frac{\ell+2}{2},\frac{\ell}{2})} 
+\Am^{\Delta}_{0,0(\frac{\ell+4}{2},\frac{\ell}{2})}\, .
\end{split}
\end{align}
We will now use this information to obtain an analytic bound on the central charge $c$ valid for any $\Nm=2$ superconformal field theory. To accomplish this, we will rely on the observation that any $\Nm=2$ SCFTs contains a closed subsector of operators isomorphic to a two-dimensional chiral algebra. Let us then start reviewing how chiral algebras appear in $\Nm=2$ SCFTs, for more details we refer the reader to the original paper \cite{Beem:2013sza} (see also \cite{Beem:2014kka,Beem:2014rza,Chester:2014mea,Lemos:2014lua}).

It is possible to define a map that associates to any $\Nm=2$ SCFTs a two-dimensional chiral algebra:
\be 
\nn
\text{$4d$ SCFT} \qquad \to \qquad \text{$2d$ Chiral Algebra}
\ee
whose correlation functions describe a protected subsector of the original four-dimensional theory. The construction of the two dimensional chiral algebra is obtained by going to the cohomology of a certain nilpotent supercharge 
\be 
\mathbb{Q} = \Qm^{1}_{-} + \bar{\Sm}^{2\, \dot{-}}\, ,
\ee
where $\Qm^{i}_{\a}$ and $\bar{\Sm}^{i\, \ad}$ are the standard supercharges of the $\Nm=2$ superconformal algebra.
Fixing a plane $\mathbb{R}^2 \in \mathbb{R}^4$ and defining complex coordinates $(z,\bar{z})$ on it, the conformal symmetry restricted to the plane acts as $SL(2) \times \overline{SL(2)}$. The supercharge $\mathbb{Q}$ can be used to define holomorphic translations that are $\mathbb{Q}$-closed and anti-holomorphic translations that are $\mathbb{Q}$-exact:
\be 
\label{sl2xhatsl2}
[\mathbb{Q}, SL(2)] = 0\, , \qquad \{\mathbb{Q},\text{something}\} = \widehat{SL(2)}\, ,
\ee
where $\widehat{SL(2)} = \text{diag}\left(\overline{SL(2)} \times SL(2)_R\right)$ and $SL(2)_R$ is the complexification of the compact $SU(2)_R$ $R$-symmetry. Operators that belong to the cohomology of $\mathbb{Q}$ transform in chiral representations of the $SL(2) \times \widehat{SL(2)}$ subalgebra. This implies that they have meromorphic OPEs (module $\mathbb{Q}$-exact terms) and their correlation functions are meromorphic functions of their positions when restricted to the plane.

In order to identify the cohomology of $\mathbb{Q}$ we will consider operators at the origin, and then we will translate them across the plane using the $SL(2) \times \widehat{SL(2)}$ generators. As shown in \cite{Beem:2013sza}, a necessary and sufficient condition for an operator to be in the cohomology of $\mathbb{Q}$ is,
\be 
\frac{1}{2}(\Delta-(j+\bar{\jmath})) - R = 0\, , \qquad r + (j-\bar{\jmath}) = 0\, .
\ee
We call this operators \textit{Schur operators}, because they contribute to the Schur limit of the superconformal index \cite{Gadde:2011uv}. It can be shown that Schur operators occupy the highest weight of their respective $SU(2)_R$ and Lorentz representations,
\be 
\Om^{1\ldots 1}_{+\ldots +\dot{+}\ldots \dot{+}}(0)\, .
\ee
Having identified the operator at the origin, we proceed to translate it using the $SL(2) \times \widehat{SL(2)}$ generators. Equation \eqref{sl2xhatsl2} implies that the anti-holomorphic dependence gets entangled with the $SU(2)_R$ structure due to the twisted nature of the $\widehat{SL(2)}$ generators. The coordinate dependence after translation is,
\be 
\Om(z,\bar{z}) = u_{i_1}(\bar{z}) \ldots u_{i_k}(\bar{z}) \Om^{(i_1 \ldots i_k)}(z,\bar{z})\, \qquad 
\text{where} \qquad u_i(\bar{z}) = (1,\bar{z})\, .
\ee
By construction, these operators define cohomology classes with meromorphic correlators. For each cohomology class we define,
\be 
\Om(z) = [\Om(z,\bar{z})]_{\mathbb{Q}}\, .
\ee
That is, to any $4d$ Schur operator there is an associated $2d$ dimensional holomorphic operator. Schur operators have protected conformal dimension and therefore sit in shortened multiplets of the superconformal algebra. In table \ref{schurTable}  we present the list of multiplets that contain a Schur operator and the holomorphic dimension $h$ of the corresponding two-dimensional operator.
\renewcommand{\arraystretch}{1.5}
\begin{table}
\centering
\begin{tabular}{|l|l|l|l|l|}
\hline
Multiplet  & $\Om_{\rm Schur}$  & $h$ & $r$    \\ 
\hline 
$\hat \Bm_R$  &  $\Psi^{11\dots 1}$   &    $R$ &  $0$ \\ 
\hline
$\Dm_{R (0, \bar{\jmath})}$  &    $ \bar{\Qm}^1_{\dot{+}} \Psi^{11\dots 1}_{\dot  + \dots \dot  + }$ &   $R+ \bar{\jmath} +1$  & $\bar{\jmath} + \frac{1}{2}$  \\
\hline
$\bar \Dm_{R (j, 0 )}$  & $ {\Qm}^1_{ +} \Psi^{11\dots 1}_{+   \dots +}$ &     $R+ j+1$  & $-j - \frac{1}{2}$  \\
\hline
$\hat \Cm_{R (j, \bar{\jmath}) }$ &   ${\Qm}^1_{+} \bar{\Qm}^1_{\dot{+}} \Psi^{11\dots 1}_{+   \dots + \, \dot  + \dots \dot  + }$&   
 $R+ j + \bar{\jmath} +2$  & $\bar{\jmath} - j$  \\
\hline
\end{tabular}
\caption{\label{schurTable}Four-dimensional superconformal multiplets that contain Schur operators, we denote the superconformal by $\Psi$. The second column indicates where in the multiplet the Schur operator sits. The third and fourth column give the two-dimensional quantum numbers in terms of $(R,j,\bar{\jmath})$.
} 
\end{table}

\subsection{Enhanced Virasoro symmetry}

Among the list of multiplets in table \ref{schurTable} is the stress-tensor multiplet $\hat{\Cm}_{0(0,0)}$ and its Schur operator is the $SU(2)_R$ conserved current $J_{+ \dot{+}}^{11}$. Its corresponding holomorphic operator is defined as $T(z)=[J_{+ \dot{+}}(z,\bar{z})]_{\mathbb{Q}}$, and the four-dimensional $J_{+ \dot{+}}(x)J_{+ \dot{+}}(0)$ OPE implies,
\be 
T(z)T(0) \sim -\frac{6\,c_{4d}}{z^4} + 2 \frac{T(0)}{z^2} + \frac{\pd T(0)}{z} + \ldots\, .
\ee
We can therefore identify $T(z)$ as the $2d$ stress-tensor. The $2d$ central charge is,
\be 
\label{c2dc4d}
c_{2d} = -12\, c_{4d}\, .
\ee
Unitarity of the four-dimensional theory implies that the two-dimensional theory is non-unitary.
The holomorphic correlator of the stress-tensor can be completely fixed in terms of the central charge, and its relation to the parent theory in four dimensions will allow us to obtain an analytic bound on $c$. The holomorphic correlator of the stress-tensor is,
\be 
\label{2dTcorrelator}
g(z) = 1 + z^4 + \frac{z^4}{(1-z)^4} + \frac{8}{c_{2d}}\left(z^2+z^3 + \frac{z^4}{(1-z)^2}+\frac{z^4}{1-z}\right)\, ,
\ee
and admits the following expansion in $SL(2)$ blocks,
\be 
g(z) = \sum_{\ell=0}^{\infty} a_{\ell}\, z^{\ell}\, {_2}F_1(\ell,\ell, 2\ell,z)\, \qquad \ell \quad \text{even,}
\ee
where ${_2}F_1$ is the standard hypergeometric function. Thanks to the $4d/2d$ correspondence we can interpret the $SL(2)$ blocks as contributions from four-dimensional multiplets containing Schur operators. Looking at the super OPE selection rules in \eqref{superOPE} there are only two possible choices,
\be 
\hat{\Cm}_{0(\frac{\ell}{2},\frac{\ell}{2})} \qquad \text{and} \qquad \hat{\Cm}_{1(\frac{\ell}{2},\frac{\ell}{2})}\, .
\ee
The $\hat{\Cm}_{0(\frac{\ell}{2},\frac{\ell}{2})}$ multiplets contain higher spin currents and we do not expect them in an interacting theory \cite{Maldacena:2011jn,Alba:2013yda}. The only candidate then is $\hat{\Cm}_{1(\frac{\ell}{2},\frac{\ell}{2})}$, the exact proportionality constant $\a$ between the OPE coefficients $\lambda^2_{\hat{\Cm}_{1(\frac{\ell}{2},\frac{\ell}{2})}}$ and the $SL(2)$ coefficients $a_{\ell}$ can be carefully worked out, but we will not need it. 
The explicit expansion of \eqref{2dTcorrelator} in terms of $SL(2)$ blocks was worked out in \cite{Beem:2015aoa}, in particular,
\be 
\lambda^2_{\hat{\Cm}_{1(\frac{1}{2},\frac{1}{2})}} = \a \left(2 - \frac{11}{15c_{4d}}\right) \, .
\ee
Unitarity of the four dimensional theory implies $\lambda^2_{\hat{\Cm}_{1(\frac{1}{2},\frac{1}{2})}} \geq 0$ then,\footnote{Because we have not calculated the exact proportionality constant, one could complain that an overall minus sign will invalidate our bound. However, common sense dictates that the sign should be positive, otherwise we will rule out every known interacting $\Nm=2$ SCFT.}
\be 
\label{analyticbound}
c_{4d} \geq \frac{11}{30}\, .
\ee
Let us note that in order to obtain this bound we only assumed $\Nm=2$ superconformal symmetry, existence of a stress-tensor, and absence of higher spin currents. Bounds of this type were obtained in \cite{Beem:2013sza} using the $\hat{\Bm}_1$ four-point function, in that case however, it is necessary to assume the existence of flavor symmetries whose conserved currents sit in $\hat{\Bm}_1$ multiplets. In the present case, our assumptions are weaker. A similar bound was also obtained for $\Nm=4$ theories in \cite{Beem:2013qxa}, where absence of higher spin currents imply $c \geq \frac{3}{4}$.

Going through the $\Nm=2$ literature one can check that the simplest rank one Argyres-Douglas fixed point (sometimes denoted as $H_0$ due to its construction in $F$-theory) has central charge $c=\frac{11}{30}$ \cite{Argyres:1995jj,Argyres:1995xn,Aharony:1998xz,Aharony:2007dj}, which precisely saturates our bound. The analytic bounds of \cite{Beem:2013sza} turned out to have interesting consequences for four-dimensional physics: the saturation of a bound was identified as a relation in the Higgs branch chiral ring due to the decoupling of the associated multiplet. It would be interesting to explore whether the absence of the $\hat{\Cm}_{1(\frac{1}{2},\frac{1}{2})}$ multiplet is associated with some intrinsic structure that characterizes the $H_0$ theory.

From the two-dimensional point of view, the $2d$ chiral algebra that describes the $H_0$ theory has been conjectured to be the Yang-Lee minimal model \cite{LR}. Indeed, the $2d$ value of the central charge is $c_{2d}=-\frac{22}{5}$. Saturation of the bound implies the absence of the $\hat{\Cm}_{1(\frac{1}{2},\frac{1}{2})}$ multiplet, from table \ref{schurTable} the associated $2d$ operator has holomorphic dimension 4. Hence, absence of $\hat{\Cm}_{1(\frac{1}{2},\frac{1}{2})}$ translate to the existence of a null state of dimension 4. Remarkably, one of the hallmarks of the Yang-Lee minimal a model is a level 4 null descendant of the identity, $\left( L_{-2}^2 - \frac{3}{5} L_{-4} \right) |0\rangle$. Our results are then consistent with the conjectured correspondence. The Schur index of Argyres-Douglas fixed points and its relation to $2d$ chiral algebras was recently studied in \cite{Buican:2015ina,Cordova:2015nma}.

The vanishing of certain OPE coefficients has also been instrumental in characterizing the $3d$ critical Ising model using numerical bootstrap techniques \cite{ElShowk:2012ht,El-Showk:2014dwa,Kos:2014bka}. One can then label the rank one $H_0$ theory as the ``Ising model'' of $\Nm=2$ superconformal theories, in the sense that it shares two of its most prominent features: minimum value of the central charge, and vanishing of certain OPE coefficients. Both features indicate that this superconformal fixed point sits in a very special place in the parameter space of $\Nm=2$ theories and a numerical treatment seems feasible \cite{Lemos:2015awa}.

\section{Superconformal block analysis}

The super selection rules are a necessary first step toward writing the conformal block expansion of the $J$ correlator. The results of section 3 give a clearer picture of how this expansion works in the case of $\Nm=2$ theories. Superconformal block expansions for $1/2$ BPS and chiral operators for several combinations of supersymmetry and spacetime dimension have been worked out \cite{Poland:2010wg,Fitzpatrick:2014oza,Dolan:2001tt,Nirschl:2004pa,Dolan:2004mu,Bobev:2015jxa}. There has been more success studying chiral and $1/2$ BPS operators because one can construct superspaces in which they are naturally defined, and the analysis simplifies. Semi-short and long multiplets in general are harder to study, the work of \cite{Khandker:2014mpa} and \cite{Doobary:2015gia} attempts to tackle the more general cases.

\subsection{Quick review of conformal blocks}

Given the four-point function of a scalar $J$ one can use the OPE in order to write the four-point correlator as a sum of conformal blocks (also called conformal partial waves),
\be 
J(x)J(0) = \sum_{\Om \in J \times J} \lambda_{J J \Om} C_{\Delta, \ell}(x,\Pm) \Om_{\Delta,\ell}(0)\, .
\ee
Plugging the OPE into the four-function,
\be 
\label{blockexp}
\langle J(x_1)J(x_2)J(x_3)J(x_4) \rangle = \frac{1}{x_{12}^4x_{34}^4} \sum_{\Om \in J \times J} \lambda_{JJ\Om}^2\,
g_{\Delta, \ell}(u,v)\, ,
\ee
where $u = \frac{x_{12}^2 x_{34}^2}{x_{13}^2 x_{24}^2}$ and $v = \frac{x_{14}^2 x_{23}^2}{x_{13}^2 x_{24}^2}$. The function $g$ is a known function of $\Delta$ and $\ell$. The dynamical information of the theory being studied is encoded in the $\Delta$s and the three-point couplings $\lambda$. The collection of $\{\Delta, \ell\}$ is called the CFT data.
The conformal blocks in four dimensions can be written explicitly in terms of hypergeometric functions \cite{Dolan:2000ut,Dolan:2003hv},
\be 
g_{\Delta, \ell}(z,\bar{z}) = \frac{z \bar{z}}{z-\bar{z}}(k_{\Delta + \ell}(z) k_{\Delta-\ell-2}(\bar{z}) - z \leftrightarrow \bar{z})\, ,
\ee
where $u=z \bar{z}$, $v=(1-z)(1-\bar{z})$, and $k_{2\beta}(z) = z^\b{_2}F_1(\b,\b,2\b,z)$. In the superconformal case a finite number of conformal families are related by supersymmetry transformations with known coefficients. This allows a rewriting of \eqref{blockexp} in terms of a ``superconformal block expansion'',
\be
\label{superblockexp}
\langle J(x_1)J(x_2)J(x_3)J(x_4) \rangle = \frac{1}{x_{12}^4x_{34}^4} \sum_{\Om \subset \Jm \times \Jm} \lambda_{\Jm \Jm \Om}^2\, \Gm_{\Delta, \ell}(u,v)\, ,
\ee
where the function $\Gm(u,v)$ is a superconformal block capturing the contributions of the superconformal multiplets appearing in \eqref{superOPE}, and it can be written as a finite sum of conformal blocks with coefficients fixed by supersymmetry

\subsection{Toward the superconformal block}

The contributions to the scalar four-point function $\langle J J J J \rangle$ are quite limited, the operators have to be $SU(2)_R \times U(1)_r$ singlets and have even spin $\ell$. We will now study the consequences of our selection rules \eqref{superOPE}. By scanning through the operator content of the different multiplets we can read which operators contribute to the expansion \eqref{superblockexp}. Below we list our findings (the ranges for $\ell$ are given in section 3).
\begin{equation}
\nn
\begin{alignedat}{4}
&{\Am_{0, 0(\frac{\ell}{2},\frac{\ell}{2})}}							&:\qquad &	g_{\Delta, \ell} + b_1\, g_{\Delta+2, \ell+2} + b_2\, g_{\Delta+2, \ell}
+ b_3\, g_{\Delta+2, \ell-2} + b_4\, g_{\Delta+4, \ell}
\qquad & \ell \quad \text{even} \\
&{\Am_{0, 0 (\frac{\ell}{2},\frac{\ell}{2})}}							&:\qquad &	g_{\Delta+1, \ell+1} + b_1\, g_{\Delta+1, \ell-1} + b_2\, g_{\Delta+3, \ell+1}
+ b_3\, g_{\Delta+3, \ell-1}
\qquad & \ell \quad \text{odd} \\
&{\Am_{0, 0 (\frac{\ell+2}{2},\frac{\ell}{2})}}							&:\qquad & g_{\Delta+2, \ell} +b_1\, g_{\Delta+2, \ell+2}	
\qquad &\ell \quad \text{even} \\
&{\Am_{0, 0 (\frac{\ell+2}{2},\frac{\ell}{2})}}							&:\qquad &	g_{\Delta+1, \ell+1} + b_1\, g_{\Delta+3, \ell+1}
\qquad & \ell \quad \text{odd} \\
&{\Am_{0, 0 (\frac{\ell+4}{2},\frac{\ell}{2})}}							&:\qquad & g_{\Delta+2, \ell+2}	
\qquad & \ell \quad \text{even} \\
&{\Cm_{0,-3(\frac{\ell}{2},\frac{\ell}{2})} }				&:\qquad &	-	\\
&{\Cm_{\frac{1}{2},-\frac{3}{2}(\frac{\ell+1}{2},\frac{\ell}{2})}}						&:\qquad & g_{6+\ell,\ell}	
\qquad & \ell \quad \text{even} \\
&{\Cm_{\frac{1}{2},-\frac{3}{2}(\frac{\ell+1}{2},\frac{\ell}{2})}}					&:\qquad & g_{7+\ell,\ell+1}	
\qquad & \ell \quad \text{odd} \\
&{\hat{\Cm}_{1(\frac{\ell}{2},\frac{\ell}{2})} }	&:\qquad &	g_{5+\ell,\ell-1} + b_1\, g_{5+\ell,\ell+1} + b_2\, g_{7+\ell,\ell+1}
\qquad & \ell \quad \text{odd}  \\
\end{alignedat}
\end{equation}
From this list we see that not all multiplets have a an associated superconformal block. Some of them contain several conformal families that contribute to the $J$ correlator, others have one single family and the associated block is just bosonic, while one multiplet does not contribute at all. For the ones that do have superconformal blocks, the $b_i$ coefficients need to be calculated. 

One way to proceed would be a brute force calculation where the $b_i$ couplings are extracted from our three-point functions. This procedure has been used for $\Nm=1$ theories and its implementation for the $\Nm=2$ case is just a straightforward generalization. However, due to the higher number of supercharges the calculation can become very cumbersome. Let us give an schematic outline of how the calculation goes for the $\Am^{\Delta}_{0,0(\frac{\ell}{2},\frac{\ell}{2})}$ block, for more details we refer the reader to \cite{Poland:2010wg}. The starting point is the superspace expansion,
\begin{align}
\label{superfield_expansion}
\begin{split}
\Om_{\a_1 \cdots \a_{\ell},\ad_1 \cdots \ad_{\ell}} & = A_{\a_1 \cdots \a_{\ell},\ad_1 \cdots \ad_{\ell}} 
+ B_{i\, \a \a_1 \cdots \a_{\ell},\ad \ad_1 \cdots \ad_{\ell}}^{\, j\,}\, \th^{\a}_{j} \thb^{\ad\, i}
\\
&
+ C_{i k\,  \a \b \a_1 \cdots \a_{\ell},\ad \bd \ad_1 \cdots \ad_{\ell}}^{\,  j l}\,
\th^{\a}_{j} \thb^{\ad\, i}\th^{\b}_{l} \thb^{\bd\, k}
+ \ldots
\end{split}
\end{align}
where $\a_1 \cdots \a_{\ell}$ and  $\ad_1 \cdots \ad_{\ell}$ are symmetrized as usual. There are also terms proportional to $(\th \thb)^3$ and $(\th \thb)^4$ that contribute to this correlator but we will ignore them to avoid cluttering. Using the superconformal algebra (see appendix A) we can write:
\begin{align}
B_{i\, \a \a_1 \cdots \a_{\ell},\ad \ad_1 \cdots \ad_{\ell}}^{\, j\,} & = \frac{1}{2} \Xi_{i\,\a\ad}^{j} A_{\a_1 \cdots \a_{\ell},\ad_1 \cdots \ad_{\ell}}\, ,
\\
C_{i k\,  \a \b \a_1 \cdots \a_{\ell},\ad \bd \ad_1 \cdots \ad_{\ell}}^{\,  j l} & = \frac{1}{16}\Xi_{i\,\a\ad}^{j} \Xi_{k\,\b\bd}^{l}
A_{\a_1 \cdots \a_{\ell},\ad_1 \cdots \ad_{\ell}}
- \frac{1}{4}\delta_{i}^{j} \delta_{k}^{l}  \Pm_{\a \ad} \Pm_{\b \bd} A_{\a_1 \cdots \a_{\ell},\ad_1 \cdots \ad_{\ell}}\, ,
\\
\vdots
\end{align}
where $\Xi_{i\,\a\ad}^{j} = [\Qm_{\a}^j,\bar{\Qm}_{i\, \ad}]$.

The next step is to build the conformal primaries associated to $B,C,\ldots$ in order to obtain the three-point couplings that relate the different conformal families inside a multiplet. Once this is accomplished we can write,
\be 
\langle J J \Om \rangle \sim \langle J J A \rangle + \lambda_{JJ B_{\text{prim}}}\langle J J B_{\text{prim}}\rangle \th \thb + \lambda_{JJ C_{\text{prim}}}\langle J J C_{\text{prim}} \rangle (\th \thb)^2 + \ldots
\ee
and the coefficients $b_i$ can be calculated from,
\be 
\frac{\lambda_{JJ B_{\text{prim}}}^2}{N_{ B_{\text{prim}}}}\, , \qquad \frac{\lambda_{JJ C_{\text{prim}}}^2}{N_{ C_{\text{prim}}}}\, , \qquad \ldots
\ee
where $N_{X} = \langle X | X \rangle$ is the norm of $X$. Although straightforward, the process becomes increasingly complicated the deeper one goes into the multiplet, i.e. the $(\th \thb)^3$ and $(\th \thb)^4$ terms.

\subsubsection{$\Nm=1$ decomposition}

Another way to organize the calculation is by splitting the $\Nm=2$ long multiplet in several $\Nm=1$ multiplets.\footnote{We are indebted to Andy Stergiou for this idea.} The idea is to organize the calculation in several $\Nm=1$ contributions and make full use of the $\Nm=1$ results already present in the literature. 
Let us start by decomposing an $\Nm=2$ multiplet in terms of $\Nm=1$ multiplets. The most efficient way to do this kind of decomposition is using superconformal characters \cite{Bianchi:2006ti,Dolan:2008qi}. The expansion works as follows, 
\begin{align}
\label{N1decomposition}
\begin{split}
\Am^{\Delta}_{0,0(\frac{\ell}{2},\frac{\ell}{2})} & \sim \Am^{\Delta}_{r_1=0(\frac{\ell}{2},\frac{\ell}{2})} + \Am^{\Delta+1}_{r_1=0(\frac{\ell-1}{2},\frac{\ell-1}{2})}
 + \Am^{\Delta+1}_{r_1=0(\frac{\ell+1}{2},\frac{\ell+1}{2})} 
\\
&
+\Am^{\Delta+1}_{r_1=0(\frac{\ell-1}{2},\frac{\ell+1}{2})} 
+\Am^{\Delta+1}_{r_1=0(\frac{\ell+1}{2},\frac{\ell-1}{2})} 
+ \Am^{\Delta+2}_{r_1=0(\frac{\ell}{2},\frac{\ell}{2})}\, ,
\end{split}
\end{align}
where $r_1 = \frac{2}{3}(r+2\,\Rm^1_{\ph{1}1})$ is the $\Nm=1$ $r$-charge after the decomposition. We have only written the $\Nm=1$ multiplets that have zero $r_1$-charge. Non-zero $r_1$-charge multiplets can not contribute to this correlator. From this expansion we conclude that only six $\Nm=1$ multiplets contribute to the OPE. Moreover, the highest dimension primary has $\Delta+2$, which means that the remaining $(\th \thb)^3$ and $(\th \thb)^4$ terms in \eqref{superfield_expansion} are $\Nm=1$ descendants, and therefore their contributions will be encoded in the $\Nm=1$ results.

The superconformal blocks for $\Nm=1$ conserved currents were worked out in \cite{Fortin:2011nq,Berkooz:2014yda,Khandker:2014mpa}. Their results read,
\begin{align}
G^{+}_{\Delta, \ell} & = g_{\Delta,\ell} +  \frac{(\Delta-2)^2(\Delta+\ell)(\Delta-\ell-2)}{16 \Delta^2 (\Delta+\ell+1)(\Delta-\ell-1)}g_{\Delta + 2,\ell}\, ,
\\
G^{-}_{\Delta, \ell} & = g_{\Delta+1,\ell-1} +  \frac{(\ell+2)^2(\Delta+\ell+1)(\Delta-\ell-2)}{\ell^2 (\Delta-\ell-1)(\Delta+\ell)}g_{\Delta + 1,\ell-1}\, , 
\end{align}
where $+(-)$ stands for $\ell$ even(odd). These results and the character identities imply that the $\Nm=2$ superconformal block can be written as,
\begin{align}
\Gm^+_{\Delta,\ell} & = G^{+}_{\Delta, \ell} + c_1\, G^{-}_{\Delta+1, \ell-1} + c_2\, G^{-}_{\Delta+1, \ell+1} + c_3\, G^{+}_{\Delta+2, \ell} + c_4 g_{\Delta+2, \ell}\, ,
\\
\Gm^-_{\Delta,\ell} & = G^{-}_{\Delta, \ell} + c_1\, G^{+}_{\Delta+1, \ell-1} + c_2\, G^{+}_{\Delta+1, \ell+1} + c_3\, G^{-}_{\Delta+2, \ell}\, .
\end{align}
The extra term in the even block represents the contributions of the $\Am^{\Delta+1}_{r_1=0(\frac{\ell-1}{2},\frac{\ell+1}{2})}$ and $\Am^{\Delta+1}_{r_1=0(\frac{\ell+1}{2},\frac{\ell-1}{2})}$ multiplets, which can not contribute to the $\ell$ odd block due to the scalar $J \times J$ selection rules. This decomposition simplifies significantly the analysis: the number of coefficients remains the same, but now we need to find primaries with dimensions up to $\Delta+2$ instead of $\Delta+4$. 

The procedure sketched above is the same, but now we have to organize the calculation in superconformal primaries annihilated by the supercharges $\Sm_1^{\beta}$ and $\bar{\Sm}^{1\, \bd}$ instead of conformal primaries annihilated by $\Km^{\bd \b}$. Once this is accomplished we can write
\be 
\langle J J \Om \rangle \sim \langle J J A \rangle + \lambda_{JJB_{\text{sprim}}}\langle J J B_{\text{sprim}}\rangle \th \thb + \lambda_{JJC_{\text{sprim}}}\langle J J C_{\text{sprim}} \rangle (\th \thb)^2\, ,
\ee
and the coefficients $c_i$ can be calculated from,
\be 
\frac{\lambda_{JJB_{\text{sprim}}}^2}{N_{B_{\text{sprim}}}}\, , \qquad \frac{\lambda_{JJC_{\text{sprim}}}^2}{N_{C_{\text{sprim}}}}\, .
\ee
This is certainly a vast simplification of the problem, but still quite involved. Also, the $\Nm=1$ decomposition \eqref{N1decomposition} works nicely for the $\Am_{R=0,r=0(\frac{\ell}{2},\frac{\ell}{2})}$ multiplet but is not so efficient for the others. For example, it is significantly more complicated for the $\hat{\Cm}_{1(\frac{\ell}{2},\frac{\ell}{2})}$ multiplet.

\section{Conclusions}

We have presented a detailed superspace analysis of all possible three-point functions of two stress-tensor multiplets
and a third arbitrary operator. From this calculation one can read the super OPE selection rules, a necessary first
step for bootstrap applications which were the main motivation behind this work.

The selection rules along with the $2d$ chiral algebra description of certain observables in $\Nm=2$ theories were sufficient
to obtain an analytic bound on $c$. This bound is valid for any $\Nm=2$ theory and is saturated by the simplest
Argyres-Douglas fixed point, denoted by $H_0$ in this paper. Saturation of the bound also implies the presence of a null state at level 4 in the associated chiral algebra, this result is consistent with the conjecture between the $H_0$ theory and the Yang-Lee minimal model.

We also presented a partial superconformal block analysis for the highest weight operator of the stress-tensor multiplet, a scalar $J$ of dimension $\Delta_J=2$. Our selection rules tell us how each conformal block organizes in a superconformal multiplet, but the
precise coefficients need to be calculated. It is not clear to us what is the most efficient way to proceed. A promising direction is
the supershadow approach developed in \cite{Khandker:2014mpa}. However, their analysis was mostly relevant for $\Nm=1$ theories in which a real operator is being exchanged. Our selection rules include $\Nm=2$ complex operators which need a generalization of their procedure. Another strategy is to perform a brute force computation, maybe the techniques used in \cite{Li:2014gpa} can be generalized to the problem at hand.

Once the superblock is obtained, the bootstrap machinery can be applied. Of great interest is the anomaly coefficient $a$, which has been absent so far from bootstrap studies. The reason is that the $a$ coefficient only participates in stress-tensor correlators, whose intricate Lorentz structure makes it a hard target for the bootstrap \cite{Dymarsky:2013wla,Echeverri:2015rwa}. One can circumvent that complication in $\Nm=2$ theories where the highest weight in the multiplet is a scalar. However, as we have seen, complications of a different nature arise when one tries to calculate the superconformal blocks. In any case, we know from the results of section 5 that what remains is just a handful of coefficients, and they should be calculable either by elegant methods or brute force. Hopefully, the ``$\Nm=2$ stress-tensor bootstrap'', which has been a glaring omission in the current numerical literature, will be a reality in the not so distant future.

\acknowledgments
We have benefited from discussions with M. Lemos, D. Li, V. Mitev, J. Park,  L. Rastelli,  D. Nandan, V. Schomerus, and A. Stergiou.    
P. L. acknowledges the hospitality of the Aspen Center for Physics, which is supported by National Science Foundation grant PHY-1066293, the Back to the Bootstrap 2015 conference at the Weizmann Institute of Science, and the Simons Summer Workshop 2015, where part of this work was performed.
P. L. is supported by SFB 647 ``Raum-Zeit-Materie. Analytische und Geometrische Strukturen''. I. R. is supported by CONICYT project No. 21120105.

\appendix

\section{$\Nm=2$ superconformal algebra}
 
In this appendix we collect our conventions for the $SU(2,2|2)$ algebra, we only list the non-vanishing commutators.

The conformal generators are $\{\Pm_{\a \ad},\, \Km^{\a \ad},\, \Mm_{\a}^{\ph{\a} \b},\, \bar{\Mm}^{\ad}_{\ph{\ad}\bd},\, D\}$. Lorentz indices transform canonically according to 
\begin{align}
&
[\mathcal{M}{_{\alpha }}^{\beta }, X_{\gamma}] = -2\delta _{\alpha }^{\beta } X_{\alpha}+\delta _{\alpha
}^{\beta } X_{\gamma}\, ,
&
&
[\mathcal{M}{_{\alpha }}^{\beta }, X^{\gamma}] = 2\delta _{\alpha }^{\gamma } X^{\b}-\delta _{\alpha
}^{\beta } X^{\gamma}\, ,
&
\\
&
[ \bar{\Mm}^{\ad}_{\ph{\ad}\bd},X_{\cd}] = 2\delta _{{\cd}}^{\ad} X_{\bd} - \delta _{\bd}^{\dot{\alpha}}X_{\cd}\, ,
&
&
[\bar{\Mm}{^{\ad}}_{\bd},X^{\cd}] = -2\delta _{{\bd}}^{\cd} X^{\ad} + \delta _{\bd}^{\dot{\alpha}} X^{\cd}\, .
&
\end{align}
The remaining $SO(4,2)$ commutators are,
\begin{equation}
[ D,\Pm_{\alpha {\dot{\alpha}}}] =\Pm_{\alpha {\dot{\alpha}}},\quad %
[ D,\Km^{\dot{\alpha}\alpha }] =-\Km^{\dot{\alpha}\alpha }
\end{equation}
and
\begin{equation}
[ \Km^{\dot{\alpha}\alpha },\Pm_{\beta \dot{\beta}}] =2\delta _{\dot{\beta%
}}^{\dot{\alpha}}\mathcal{M}{_{\beta }}^{\alpha }-2\delta _{\beta }^{\alpha }%
\bar{\mathcal{M}}{^{\dot{\alpha}}}_{\dot{\beta}} -4\delta
_{\beta }^{\alpha }\delta _{\dot{\beta}}^{\dot{\alpha}}D\, .
\end{equation}
The $SU(2)_R \times U(1)_r$ $R$-symmetry generators are denoted by $\{\Rm{^{i}}_{j}, \, r \}$. $SU(2)_R$ indices transform according to,
\be
[\Rm{^{i}}_{j}, X_{k}] = -\delta _{k}^{i} X_{j}+\frac{1}{2}\delta _{j}^{i} X_{k}\, ,
\qquad
[\Rm{^{i}}_{j}, X^{k}] = \delta _{j}^{k} X^{i}-\frac{1}{2}\delta _{j}^{i} X^{k}\, .
\ee
The fermionic generators are the Poincar\'e and conformal supercharges are $\{\Qm^i_{\a},\bar{\Qm}_{i\, \ad}, \Sm_{i}^{\a},\bar{\Sm}^{i\, \ad}\}$ and their anticommutators are,
\begin{eqnarray}
\{\Qm_{\alpha }^{i},\bar{\Qm}_{j\,{\dot{\alpha}}} \} &=&-2\mathrm{i}\delta
_{j}^{i} \Pm_{\alpha {\dot{\alpha}}},\quad \{ \bar{\mathcal{\Sm}}{^{i\,\dot{%
\alpha}},\Sm}_{j}^{\alpha } \} =2\mathrm{i}\delta _{j}^{i}\Km^{\dot{\alpha}\alpha }
\\
\{\Qm_{\alpha }^{i},{\Sm}_{j}^{\beta } \} &=&-2\delta _{j}^{i}%
\mathcal{M}{_{\alpha }}^{\beta }+2\delta _{j}^{i}\delta _{\alpha }^{\beta
}\left( D-r\right) -4\delta _{\alpha }^{\beta }\Rm{^{i}}_{j} \\
\{ \bar{\mathcal{\Sm}}{^{i\,\dot{\alpha}},}\bar{\Qm}_{j\,\dot{\beta}} \}
&=&-2\delta _{j}^{i}\bar{\mathcal{M}}{^{\dot{\alpha}}}_{\dot{\beta}}-2\delta
_{j}^{i}\delta _{\dot{\beta}}^{\dot{\alpha}}\left( D+r\right) -4\delta _{%
\dot{\beta}}^{\dot{\alpha}}\Rm{^{i}}_{j}\, .
\end{eqnarray}%
Finally, the commutators between bosonic and fermion generators,
\begin{align}
[ \Km^{\dot{\beta}\beta },\Qm_{\alpha }^{i} ] &= -2\mathrm{i}\delta _{\alpha
}^{\beta }\bar{\Sm}{^{i\,\dot{\beta}}},\quad [ \Km^{\dot{\beta}%
\beta },\bar{\Qm}_{i\,{\dot{\alpha}}}] =2\mathrm{i}\delta _{\dot{\alpha}}^{\dot{%
\beta}}{S}_{i}^{\beta }\, , \\
[ \Pm_{\beta \dot{\beta}},{S}_{i}^{\alpha }] &= -2\mathrm{i}\delta _{\beta
}^{\alpha }\bar{Q}_{i\,\dot{\beta}},
\quad [ \Pm_{\beta \dot{\beta}},\bar{%
\Sm}{^{i\,\dot{\alpha}}}] {=2\mathrm{i}}\delta _{\dot{\beta}}^{\dot{\alpha%
}}\Qm_{\beta }^{i}\, ,
\end{align}
and
\begin{align}
&
[ D,\Qm_{\alpha }^{i} ] = \frac{1}{2}\Qm_{\alpha }^{i}\, ,
&
&
[ D, \Sm_{i}^{\alpha } ] =-\frac{1}{2}\Sm_{i}^{\alpha }\, ,
&
&
[D,\bar{\Qm}_{i\,{\dot{\alpha}}} ] = \frac{1}{2}\bar{\Qm}_{i\,{\dot{%
\alpha}}}\, ,
&
&
[D,\bar{\Sm}{^{i\,\dot{\alpha}}} ] = - \frac{1}{2}\bar{\Sm}{^{i\,\dot{\alpha}}}\, ,
&
\\
&
\, \, \, [r,\Qm_{\alpha }^{i} ] =\frac{1}{2}\Qm_{\alpha }^{i}\, ,
&
&
\, \, \,[r, \Sm_{i}^{\alpha } ] =-\frac{1}{2}\Sm_{i}^{\alpha }\, ,
&
&
\, \, \, [r,\bar{\Qm}_{i\,{\dot{\alpha}}} ] = -\frac{1}{2}\bar{\Qm}_{i\,{\dot{%
\alpha}}}\, ,
&
&
\, \, \, [r,\bar{\Sm}{^{i\,\dot{\alpha}}} ] = \frac{1}{2}\bar{\Sm}{^{i\,\dot{\alpha}}}\, .
&
\end{align}

\section{Superspace identities}

Here we collect some superspace identities necessary for the three-point functions calculations of Section 3. Let us start proving that if equations \eqref{dTheta2} and the $\mathbb{Z}$ symmetry condition are satisfied, then equations \eqref{D2} are also satisfied. In general the function $H$ will always be expressible as sum of the form
\be 
H(\mathbf{Z}_3) = \sum_a f_a(\Xbf_3)g_a(\Theta_3,\bar{\Theta}_3)\, ,
\ee 
where the functions $g_a$ satisfy the conditions,
\be 
\frac{\pd^2}{\pd \Theta^{i}_{3\,\a} \pd \Theta^{\a\, j}_{3}}\, g_a(\Theta_3, \bar{\Theta}_3) = 0\, ,
\qquad
\frac{\pd^2}{\pd \bar{\Theta}^{\ad}_{3\,i} \pd \bar{\Theta}}_{3\,\ad\, j}\, g_a(\Theta_3, \bar{\Theta}_3) = 0\, ,
\ee
and $g_a(-\Theta_3,-\bar{\Theta}_3) = (-1)^F g_a(\Theta_3,\bar{\Theta}_3)$ where $(-1)^F$ counts the number of $\Theta$s and $\bar{\Theta}$s. From \eqref{Ds} it follows,
\be 
\Dm_{\a\, i}\, \bar{\Xbf}_{3\, \a \ad} = 0\, , \qquad \tilde{\Dm}^{i}_{\ad}\, \bar{\Xbf}_{3\, \a \ad} = 0\, ,
\ee
and,
\begin{align}
\begin{split}
\Dm_{\a\, i} \Dm^{\a}_{j}\, H(\Xbf_3, \Theta_3, \bar{\Theta}_3) & = \Dm_{\a\, i} \Dm^{\a}_{j} \sum_{a} f_a(\Xbf_3) \, g_a(\Theta_3,\bar{\Theta}_3)
\\
& = \sum_{a} f_a(-\bar{\Xbf}_3)\Dm_{\a\, i} \Dm^{\a}_{j} \, g_a(-\Theta_3,-\bar{\Theta}_3)
\\
& = \sum_{a} f_a(-\bar{\Xbf}_3) (-1)^F \frac{\pd^2}{\pd \Theta^{i}_{3\,\a} \pd \Theta^{\a\, j}_{3}}\, g_a(\Theta_3, \bar{\Theta}_3)
\\
& = 0
\end{split}
\end{align}
as promised. Thanks to this property imposing the conservation constraint is now an algebraic exercise, Taylor expanding both sides of the $\mathbb{Z}_2$ equation and equating coefficients solves the problem. For the expansion of the denominators we use the following identity,
\begin{align}
\nn
\frac{1}{(\bar{\Xbf}^2)^{\Delta}} & =  \frac{1}{(\Xbf^2)^{\Delta}} -4\mathrm{i}\,\Delta \frac{(\Theta^{\alpha i}\Xbf_{\alpha \dot{\alpha}}\bar{\Theta}^{\dot{\alpha}}_i)}{(\Xbf^2)^{\Delta+1}}
-8\,\Delta(\Delta-1)  \frac{(\Theta^{\alpha i}\Xbf_{\alpha \dot{\alpha}}\bar{\Theta}^{\dot{\alpha}}_i)^2}{(\Xbf^2)^{\Delta+2}}
-8\,\Delta \frac{(\Theta^{\alpha \beta}\Xbf_{\alpha \dot{\alpha}}\Xbf_{\beta \dot{\beta}}\bar{\Theta}^{\dot{\alpha} \dot{\beta}})}{(\Xbf^2)^{\Delta+2}} 
\\
& 
+\frac{32}{3}\mathrm{i}\,\Delta(\Delta^2-1) \frac{(\Theta^{\alpha i}\Xbf_{\alpha \dot{\alpha}}\bar{\Theta}^{\dot{\alpha}}_i)^3}{(\Xbf^2)^{\Delta+3}}
+\frac{32}{3}\,\Delta^2(\Delta^2-1) \frac{(\Theta^{\alpha i}\Xbf_{\alpha \dot{\alpha}}\bar{\Theta}^{\dot{\alpha}}_i)^4}{(\Xbf^2)^{\Delta+4}}\, ,
\end{align}
which for the special case $\Delta=1$ becomes eq. (3.27) in \cite{Kuzenko:1999pi}. After Taylor expanding what remains is to write all terms in our equation using the same basis of Grassmann objects. As usual with fermions, high powers of Grassmann variables are not all independent, for example,
\be 
(\Theta^{\alpha i}\Xbf_{\alpha \dot{\alpha}}\bar{\Theta}^{\dot{\alpha}}_i)^3 = (\Theta^{\alpha \beta}\Xbf_{\alpha \dot{\alpha}}\Xbf_{\beta \dot{\beta}}\bar{\Theta}^{\dot{\alpha} \dot{\beta}})(\Theta^{\c\, i}\Xbf_{\c \cd}\bar{\Theta}^{\cd}_i) \, .
\ee
Several identities of this type were needed for the calculations of section 3, we implemented them in Mathematica using the grassmannOps.m package by J. Michelson and M. Headrick.


\bibliographystyle{JHEP}
\bibliography{bibliography}{}

\end{document}